\documentclass[useAMS]{mn2e}
\usepackage{epsfig,natbib_vw,times}

\usepackage{amssymb,amsmath}

\bibpunct[, ]{(}{)}{;}{a}{}{,}

\def \aj {AJ}
\def \mnras {MNRAS}
\def \apj {ApJ}
\def \apjs {ApJS}
\def \apjl {ApJL}
\def \aap {A\&A}

\def \araa {ARAA}

\def \zabs {z_{\rm abs}}
\def \zqso {z_{\rm qso}}
\def \caii {Ca~{\sc ii}}
\def \caiii {Ca~{\sc iii}}

\def \mgii {Mg~{\sc ii}}
\def \feii {Fe~{\sc ii}}
\def \mgi {Mg~{\sc i}}

\def \hi {H~{\sc i}}
\def \nhi {$N$(H~{\sc i})}

\def \EBV {\textit{E(B$-$V)}}
\def \oii {[O~{\sc ii}]}
\def \kms {\rm km\,s^{-1}}
\def \sfrd {\dot{\rho}^*}
\def \ergscm {\rm ergs\,cm^{-2}\,s^{-1}}
\def \Msolyr {\rm M_\odot yr^{-1}}
\def \perkpcsq {\rm kpc^{-2}}
\def \percmsq {{\rm cm}^{-2}}

\def \be {\begin{equation}}
\def \ee {\end{equation}}

\def\gsim{\mathrel{\lower0.6ex\hbox{$\buildrel {\textstyle >}
 \over {\scriptstyle \sim}$}}}
\def\lsim{\mathrel{\lower0.6ex\hbox{$\buildrel {\textstyle <}
 \over {\scriptstyle \sim}$}}}

\def\m@th{\mathsurround=0pt }
\def\eqalign#1{\null\,\vcenter{\openup1\jot \m@th
 \ialign{\strut\hfil$\displaystyle{##}$&$\displaystyle{{}##}$\hfil
 \crcr#1\crcr}}\,}

\title[The star-formation rate of DLAs at $z \lsim 1$]{The star-formation rate
of \caii\ and damped Lyman-$\alpha$ absorbers at $0.4<z<1.3$}
\author[V. Wild et~al.]{Vivienne
  Wild$^1$\thanks{vwild@mpa-garching.mpg.de}, Paul C. Hewett$^2$ and Max Pettini$^2$
\vspace*{6pt}\\
$^1$Max Planck Institute for Astrophysics, Karl-Schwarzschild Str. 1,
85748 Garching, Germany \\
$^2$Institute of Astronomy, University of Cambridge, Madingley Road,
Cambridge CB3 0HA, UK\\}

\begin{document}

\maketitle

\begin{abstract}

Using stacked Sloan Digital Sky Survey spectra, we present the
detection of \oii\,$\lambda\lambda$3727,3730 nebular emission from the
galaxies hosting \caii\ absorption line systems and galaxies hosting
\mgii-selected damped Lyman-$\alpha$ (DLA) absorbers.  Both samples of
absorbers, 345 \caii\ systems and 3461 \mgii-selected DLA systems,
span the redshift interval $0.4 \le \zabs < 1.3$; all of the former
and half the latter sample are expected to be bona-fide DLAs.  The
measured star formation rate (SFR) per absorber from light falling
within the SDSS fibre apertures (corresponding to physical radii of
$6-9 h^{-1}\,$kpc) is $0.11-0.14\,\Msolyr$ for the \mgii-selected DLAs
and $0.11-0.48\,\Msolyr$ for the \caii-absorbers.  These results
represent the first estimates of the average SFR in an
absorption-selected galaxy population from the direct detection of
nebular emission. Adopting the currently favoured model in which DLAs
are large, with radii $\gsim 9h^{-1}\,$kpc, and assuming no
attenuation of the \oii\ emission by dust, leads us to conclude that
the SFR per unit area of \mgii-selected DLAs falls an order of
magnitude below the predictions of the Schmidt law, which relates the
SFR to the \hi\ column density at $z$$\sim$$0$. While DLA sightlines
are known to contain little dust, the unknown geometry of the dust
distribution in the galaxies causes the main uncertainty in our results.

The contribution of both DLA and \caii\ absorbers to the total
observed star formation rate density, $\sfrd$, in the redshift range
$0.4 < z < 1.3$, is small, $\lsim10\%$ and $\lsim3\%$ respectively.
The result contrasts with the conclusions of Hopkins et al. that DLA
absorbers can account for the majority of the total observed $\sfrd$
in the same redshift range. The disagreement is a direct consequence
of the much lower SFR per unit area we observe than predicted by the
Schmidt law.  Our results effectively rule out a picture in which DLA
absorbers are the sites in which a large fraction of the total $\sfrd$
at redshifts $z \lsim 1$ occurs.

\end{abstract}


\section{Introduction}\label{sec:intro}

Uncovering the physical properties of the host galaxies of strong
quasar absorption line systems is a key topic in extragalactic
astronomy today.  These absorbers are expected to select galaxies
based on gas cross-section, rather than luminosity, providing a
complementary view of the galaxy population to that obtained from
traditional stellar luminosity selected galaxy samples.  However, with
the possible exception of the common \mgii\,$\lambda\lambda$2796,2803
absorption line systems, the identification of large samples of
absorption-selected objects with known emission properties has not
been possible, despite considerable effort. A complete picture
relating the gas cross-section, ionisation state and chemical
abundances of the interstellar media and gaseous halos surrounding
galaxies to the emission properties of the associated galaxy remains
elusive.

At high redshift ($z$$\gsim$2), observational efforts have focused on
the highest neutral hydrogen (\hi) column density systems, Damped
Lyman-$\alpha$ systems (DLAs).  These systems contain the majority of
neutral gas at high redshifts and may provide the fuel
reservoirs for subsequent star formation in galaxies
\citep{1995ApJ...440..435L}.  Emission, in general from Lyman-$\alpha$,
from less than a dozen host galaxies of high redshift DLAs has been
detected and spectroscopically confirmed: a summary is given in
\citet{2005MNRAS.358..985W}.  Adopting only modest corrections to
allow for the presence of dust, both detections and limits to the
inferred star-formation rates (SFR) of these systems are relatively large,
$\gsim$10$\,\Msolyr$.

The small, inhomogeneous samples of counterpart galaxies have
precluded direct estimates of the total SFR density ($\sfrd$) of
galaxies associated with high redshift absorption systems; however,
two methods have been used to obtain indirect estimates.
\citet{2005ApJ...630..108H} extrapolated the low redshift relation
between \hi\ column density and SFR to high redshift; while Wolfe et
al. (2003a, 2003b) \nocite{2003ApJ...593..215W, 2003ApJ...593..235W}
focused on the use of absorption from the fine-structure levels of the
ground state of C~{\sc ii} to estimate the SFR per unit area in DLAs
and to calculate their contribution to $\sfrd$ at high redshift.  Both
approaches involve significant assumptions and the conclusions of the
two studies are only marginally consistent, with
\citet{2003ApJ...593..235W} advocating a larger contribution to the
SFR density than \citet{2005ApJ...630..108H}.

At lower redshifts ($z$$\lsim$$1$), galaxies associated with
absorption-line systems are more readily observed because the angular
separation between the galaxy and quasar is greater and deep imaging
allows the detection of intrinsically fainter galaxies.  Deep imaging
and follow-up spectroscopic surveys of the fields of both DLAs and
\mgii\ absorbers have revealed host galaxies with a wide range of
morphological types and at varying impact parameters from the
background quasar \citep{1991A&A...243..344B, 1993A&A...279...33L,
1994ApJ...437L..75S, 1997ApJ...480..568S, 1997A&A...321..733L,
2003ApJ...595...94R, 2005ApJ...620..703C}. Small samples, with
incomplete spectroscopic follow-up, again hinder inference of the SFR
density of absorption-selected galaxies. For a recent review on the
nature of galaxies associated with \mgii\ absorbers see
\citet{2005ASPC..331..387C}.

Recently, \citet{2005MNRAS.361L..30W} identified a new class of quasar
absorption line system selected by their strong \caii\,
$\lambda\lambda3935,3970$ absorption doublet\footnote{Vacuum
wavelengths are used throughout this paper.}  from quasar spectra in
the Sloan Digital Sky Survey (SDSS).  These rare (number densities
$\sim 30$\% of DLAs) absorption line systems contain significant
quantities of dust -- $\langle\EBV\rangle$$\sim$$0.1$ -- in contrast
to other samples of quasar absorption line systems
\citep{1997ApJ...478..536P, 2004MNRAS.354L..31M, 2006MNRAS.tmp..274Y}.
The dust and metal line properties of the \caii\ absorbers strongly
suggests they have \hi\ column densities greater than the nominal
limit for DLAs of $10^{20.3}$\,atoms\,$\percmsq$ \citep[][hereafter
WHP06]{2006MNRAS.367..211W}.


The presence of strong \caii\ absorption in absorption line systems is
of particular interest for studying the properties of the interstellar
medium (ISM) of galaxies.  Due to the convenient position of its
resonance lines in the optical spectrum, \caii\ has been well studied
in the disk of the Milky Way (MW): its distribution is found to vary
smoothly within the warm neutral medium and extend to a scale height
of about $1$\,kpc from the plane of the disk
\citep{1989ApJ...340..762E, 1996ApJS..106..533W, 1997ApJS..112..507W,
astro-ph/0601363}.  Calcium is severely depleted onto dust grains in
the MW \citep{1996ARA&A..34..279S} and, due to its second ionisation
potential of 11.87eV being below the ionisation potential of hydrogen,
\caii\ is a trace ionisation state in the ISM of
galaxies -- most calcium is in the form of \caiii.  The smoothness of the
distribution, despite the sensitivity of gas phase \caii\ to ionising
radiation fields and dust, have led to the suggestion that it traces
warm neutral gas clouds within the MW, along with similarly depleted
elements such as titanium \citep{astro-ph/0601363}.  Alternative
explanations for the detection of strong \caii\ lines in absorbers at
$z$$\sim$$1$ could be the destruction of dust grains, perhaps by shocks
releasing calcium into the gas-phase, or lower radiation fields, perhaps
within high volume density clouds which afford a degree of self
shielding.


WHP06 investigated the dust content of the absorbers by measuring both
the reddening of the background quasar and dust depletion patterns of
different ions detected in stacked SDSS spectra.  A dust-to-metals
ratio similar to that of the MW was found, suggesting a higher degree
of chemical evolution than in other quasar absorption line systems
\citep{2004A&A...421..479V}.  The extent of depletion of refractory
elements in the absorbers with highest equivalent width (EW) of the
\caii\,$\lambda$3935 line ({$W_{\lambda3935}$) was shown to equal that
seen in the warm neutral medium of the MW, in contrast to average
DLAs.

In this paper we present the detection of the
\oii\,$\lambda\lambda$3727,3730 emission doublet\,\footnote{Component
vacuum wavelengths of 3727.09\,\AA\ and 3729.88\,\AA.}  associated
with \caii\ absorbers.  The detection of \oii\ confirms the presence of
nearby star formation and allows, for the first time, a direct
estimate of the average SFR in an absorption-selected galaxy
population.  In order to provide a comparison to a more commonly
studied sample of absorption line systems, we also analyse a sample of
strong \mgii-selected absorbers in which 50\% are expected to be DLAs.
Again, \oii\,$\lambda\lambda$3727,3730 emission is detected.

The outline of this paper is as follows.  In Section \ref{sample} we
review the sample selection, extending our original \caii\ and \mgii\
absorber samples to lower redshifts and lower signal-to-noise ratio
threshold.  We also detail the methods used to create the composite
spectra and measure the \oii\ emission line luminosities.  In Section
\ref{sfr} we present the measured \oii\ line luminosities and discuss
the magnitude of corrections required to convert these measurements
into a total SFR for the absorber host galaxies.  We calculate lower
limits on the SFR per unit area of the absorbers, based simply on our
observations and the size of the SDSS fibre aperture. Estimates of the
SFR per unit area and the contribution of the absorber host galaxies
to the volume averaged SFR density ($\sfrd$) are derived in Section
\ref{abssfrprop} which also includes a discussion of the combined
uncertainties involved in these measurements. Finally, Section
\ref{discussion} discusses the implications of the observational
results for the applicability of the Schmidt law (Schmidt 1959),
relating the gas surface density to the SFR, to the absorbers and how
our results compare with other recent work.  Unless otherwise stated,
a flat cosmology with $\Omega_\Lambda=0.7$, $\Omega_M=0.3$,
$H_0=100\,h\,\kms\,{\rm Mpc}^{-1}$ and, where necessary, $h=0.7$ is
assumed throughout the paper.

\section{Method}\label{sample}

Quasar spectra were selected from the SDSS DR3 catalogue of
\citet{2005AJ....130..367S} and supplemented with spectra in Data
Release 4 \citep[DR4,][]{2006ApJS..162...38A} main survey plates and
spectroscopically classified as a quasar or high-redshift
quasar\,\footnote{'SpecClass' equal to $3$ or $4$ in an SQL search of
the DR4 catalogue}.  Quasars in the sample were restricted to those
with extinction-corrected point spread function (PSF) $i$-band
magnitude ($m_{psf,i}$) brighter than $19.1$.  The redshift range of
the quasars was restricted to $0.41<\zqso<3.1$; the lower limit set to
ensure that the presence of the \mgii\,$\lambda\lambda2796,2803$
doublet could be used to confirm the \caii\ detections.  To ensure our
analysis was confined to relatively high quality spectra, the quasar
spectra were required to possess $\ge3700$ good pixels ({\sc
``NGOOD''}) and spectral signal-to-noise ratio (SNR) $\ge 6.0$ in both
``i'' and ``r'' bands \footnote{A further $\sim 150$ spectra suffering
from a variety of artifacts, including many spectra in spectroscopic
plates 426 and 946 included in the DR3 release, were removed from the
sample following visual inspection.}.

Quasar spectra showing evidence for Broad Absorption Line (BAL) and
strong associated absorption features were removed from the sample.
This was achieved using a preliminary Principal Component based
identification scheme similar to that of WHP06, followed by
confirmation through visual inspection.  Comparison of our BAL and
strong absorber catalogue to that of \citet{astro-ph/0603070} shows
the two agree extremely well; it was not possible to simply adopt the
\citet{astro-ph/0603070} DR3 catalogue because our sample includes
quasars in the DR4 release.  While it is important to remove the
`BAL'-quasars from our sample to obtain accurate dust reddening
estimates, none of the results described in the paper are sensitive to
the exact identification scheme used.  In the DR3 and DR4 samples of
27\,010 and 7653 quasar spectra, 1812 and 573 objects respectively
were identified as possessing BALs or strong associated absorption,
leaving a combined sample of 32\,278 quasar spectra in which
absorption systems were sought.

\subsection{The absorber sample}\label{sample_detail}

A matched-filter search was employed to detect
\mgii\,$\lambda\lambda2796,2803$ absorption doublets with SNR$\ge$6
over the redshift interval 0.4$<\zabs<$1.3 in the quasar
spectra\footnote{The SNR is calculated from the combined detection of
both members of the doublet.}.  The search for \mgii\ doublets was
restricted such that the absorbers appeared at wavelengths
$>$1250\,\AA\ in the quasar rest frame, i.e.  not in the
Lyman-$\alpha$ forest, and had redshifts $\zabs<\zqso-0.03$.  The
quasar spectra were then searched independently for
\caii\,$\lambda\lambda$3935,3970 doublet absorbers with rest frame
$W_{\lambda3935} >0.3$\AA\ and SNR$\ge$ 4 over the redshift interval
$0.4<\zabs<1.3$.  The impact of imperfect removal of the OH skylines
in the red half of the SDSS spectra was minimised by employing the
routine of \citet{2005MNRAS.358.1083W}, with wavelength regions
potentially occupied by known absorption lines masked.  Further
details of the line-finding procedure are given in WHP06.

The distribution of $\zabs$(\mgii)$-\zabs$(\caii) shows a very strong
excess of absorbers centred on zero-velocity difference with a
dispersion of $70\,{\rm km s}^{-1}$.  An initial selection of 360
candidate \caii\ absorbers was defined by selecting \caii\ absorbers
within a velocity difference of $\pm200\,{\rm km s}^{-1}$ of a
detected \mgii\ absorber.  The fraction of spurious \caii\ absorbers
can be estimated directly from the incidence of apparent \caii\
absorbers with velocity differences larger than $\pm200\,{\rm km
s}^{-1}$.  The level of contamination is well-determined and only 5\%
of the sample of 360 candidate absorbers are predicted to be due to
chance coincidence.

Absorption line properties of the \feii\,~$\lambda$2600,
\mgii\,$\lambda\lambda$2796,2803 and \mgi\,$\lambda$2853 features for
all of the \mgii\ absorber candidates were calculated via a fully
parameterised fit to the spectra\,\footnote{MPFIT,
{http://cow.physics.wisc.edu/$\sim$craigm/idl/}, Craig Markwardt IDL
library}.  The \caii\ doublet was included for the subsample of \caii\
absorbers.  A continuum was fitted to the regions around the \mgii\
and \caii\ lines and the corresponding portions of the quasar spectra
were normalised by dividing by the continuum level.  Gaussian doublets
were then fitted to the normalised spectra using a maximum-likelihood
routine; the position and line width of the doublet, and the relative
strengths of the individual lines, were allowed to vary freely.  Rest
frame equivalent widths ($W$) were calculated from the parameters of
the Gaussian fits and errors estimated by propagation of the parameter
errors derived during the maximum likelihood fit.

Each candidate \caii\ absorption line system was inspected visually and 15
spurious detections were removed.  Following \citet{2006ApJ...636..610R}, \mgii\
systems were restricted to those with $1<W_{\lambda2796}/W_{\lambda2600}<2$ and
$W_{\lambda2796}>0.6$ and only those systems with \feii\,$\lambda$2600 and
\mgii\,$\lambda$2796 lines detected with a SNR$\ge$3 (see Equation
\ref{eq:snr}).  This sample, referred to from now on as \mgii-selected DLA
candidates, was selected to maximise the fraction of DLA systems it contains,
about 50\%, thus providing a comparison to previous results for DLAs discussed
in Section \ref{sec:intro}, and to the \caii\ systems which are expected to be
DLAs.  Our final absorber catalogues contain 3461 \mgii\ absorption line systems
and 345 \caii\ absorbers in the redshift range $0.4\le\zabs\le1.3$.

\begin{figure}
  \begin{minipage}{\textwidth}
    \includegraphics[scale=0.7]{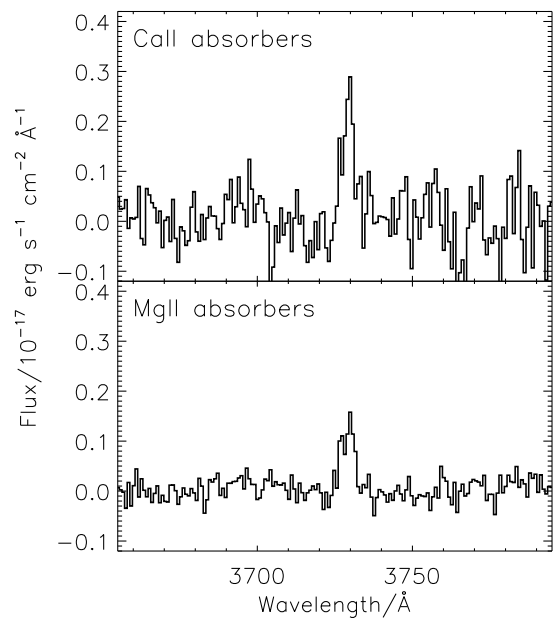}
  \end{minipage}
\caption{The wavelength region $3655-3795\,$\AA\ of a composite of
the 345 \caii\ absorbers (Upper Panel) and of a composite of 3461
\mgii\ absorbers (Lower Panel) in the absorber rest frame. The
individual spectra have simply been moved to the absorber rest frame,
normalised and averaged together.}
\label{base_comp}
\end{figure}

The samples are more extensive than those reported in WHP06 due to (i) the
increased redshift interval searched, $\zabs>0.4$ compared to $\zabs>0.84$, (ii)
the larger number of quasar spectra used by reducing the spectrum SNR threshold
from $10.0$ to $6.0$, and (iii) the lower significance threshold for the
detection of the \caii\ absorbers, reduced from SNR$>5.0$ to SNR$>4.0$, subject
to the presence of \mgii\ absorption within $\pm200\,{\rm km s}^{-1}$.

Before considering the detailed procedures employed to obtain
measurements of the \oii\,$\lambda\lambda$3727,3730 emission
associated with the samples, it is worth stressing the potential of
the SDSS spectroscopic database for such investigations.  The typical
1$\sigma$ noise in the SDSS quasar spectra is $\sim 1 \times
10^{-17}\ergscm$ per pixel ($\Delta v = 69\kms$), and detection limits
for (close to) unresolved features in several hundred or more stacked
spectra can easily reach $10^{-18}\ergscm$.  At redshifts of $z$$\sim$1
such fluxes translate to unattenuated \oii\ emission corresponding to
SFRs of significantly below 1$\,\Msolyr$.  Figure \ref{base_comp}
shows the wavelength region $3655-3795\,$\AA\ of composites of the
\caii\ and \mgii-selected DLA absorbers in the absorber rest frame; the
individual spectra have simply been moved to the absorber rest frame,
normalised and averaged together.  The clear presence of the \oii\
emission line is striking, including the partial resolution of the
doublet.

\subsection{Building the composite spectra}

For the quantitative investigation of the \oii\ emission line
properties of the absorbers, the spectra were first corrected for the
effects of Galactic extinction, then moved to the absorber rest frame,
without rebinning, correcting the flux into rest-frame per-\AA\ units.
Continua in the region of \oii\ were defined using a running median
filter (size 61 pixels) and subtracted from each spectrum.  The
residual spectra were weighted by the square of the luminosity
distance to the absorber and combined into a composite using an
arithmetic mean
\footnote{In this situation it is not possible to use a weighted mean
due to the correlation between absorber redshift and SNR of the
spectra: low-redshift absorbers can be detected in lower-redshift
quasars which have generally higher SNR spectra.}.

The absorber sample covers an extensive redshift interval, over which
it is believed the average SFR in galaxies changed significantly
\citep{2004ApJ...615..209H}.  Composites were therefore created in two
redshift bins, $\zabs=0.4-0.8$ and $\zabs=0.8-1.3$.  For the \caii\
absorbers these redshift subsets were further split into two by
$W_{\lambda3935}$, because of the known variation of dust properties
with strength of \caii\ absorption \citep{2005MNRAS.361L..30W}.  An
EW-limit of $W_{\lambda3935} = 0.68$\,\AA\ was chosen for this
division to match that used in WHP06 but the results reported below
are not sensitive to the exact value of EW used.  Unlike any of the
\caii\ absorbers, $\sim 0.2\%$ of the \mgii\ absorber spectra show
detectable \oii\ emission in individual spectra.  However, the
contribution to the mean \oii\ luminosity of the absorber sample by
these objects is small and the results of the paper are not dependent
on the inclusion of the small percentage of such systems.  The
composite spectra in the region of the \oii\ line are shown in Figs.
\ref{sfr_ca_mg} and \ref{sfr_ca}.

\begin{figure*}
  \begin{minipage}{\textwidth}
    \includegraphics[scale=0.8]{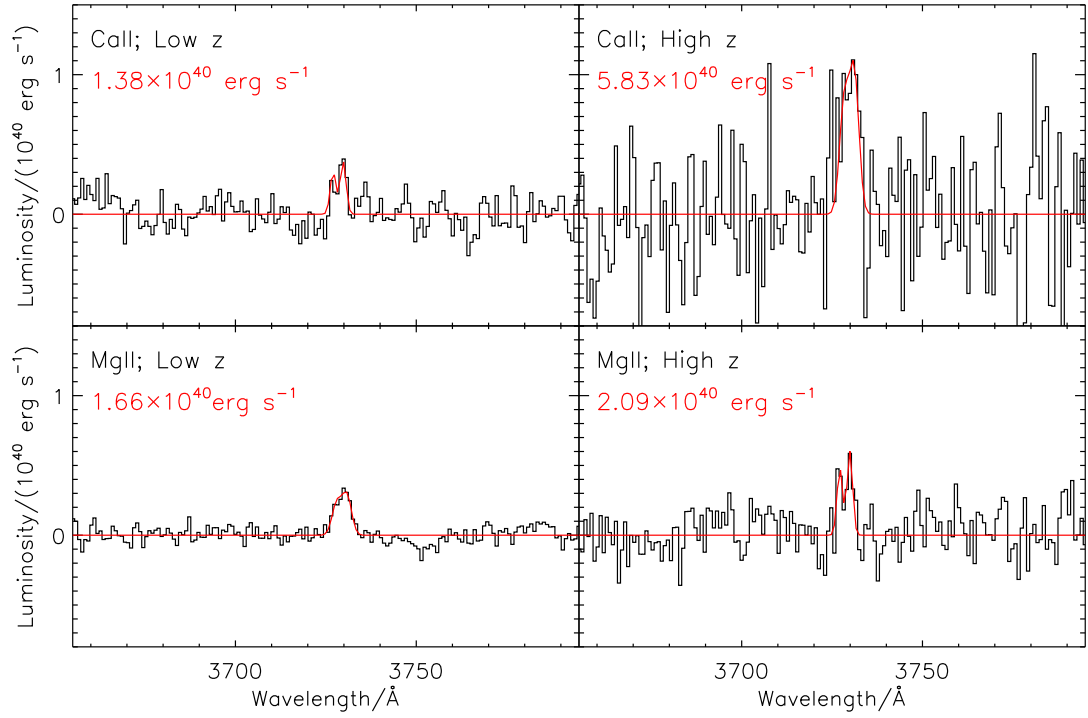}
  \end{minipage}
  \caption{SDSS composite spectra of \caii\ absorption line systems
  (top) and \mgii-selected DLA candidates (bottom)
  in the region of \oii\,$\lambda\lambda$3727,3730 in
  redshift bins of $0.4 \le \zabs < 0.8$ (left) and $0.8 \le \zabs < 1.3$
  (right). Flux from each contributing spectrum has been converted
  into luminosity before combining. Overplotted (in red lines)
  are Gaussian fits to the emission lines. Measured line luminosities
  from these line fits are given in each plot.}
  \label{sfr_ca_mg}
\end{figure*}

\begin{figure*}
  \begin{minipage}{\textwidth}
    \includegraphics[scale=0.8]{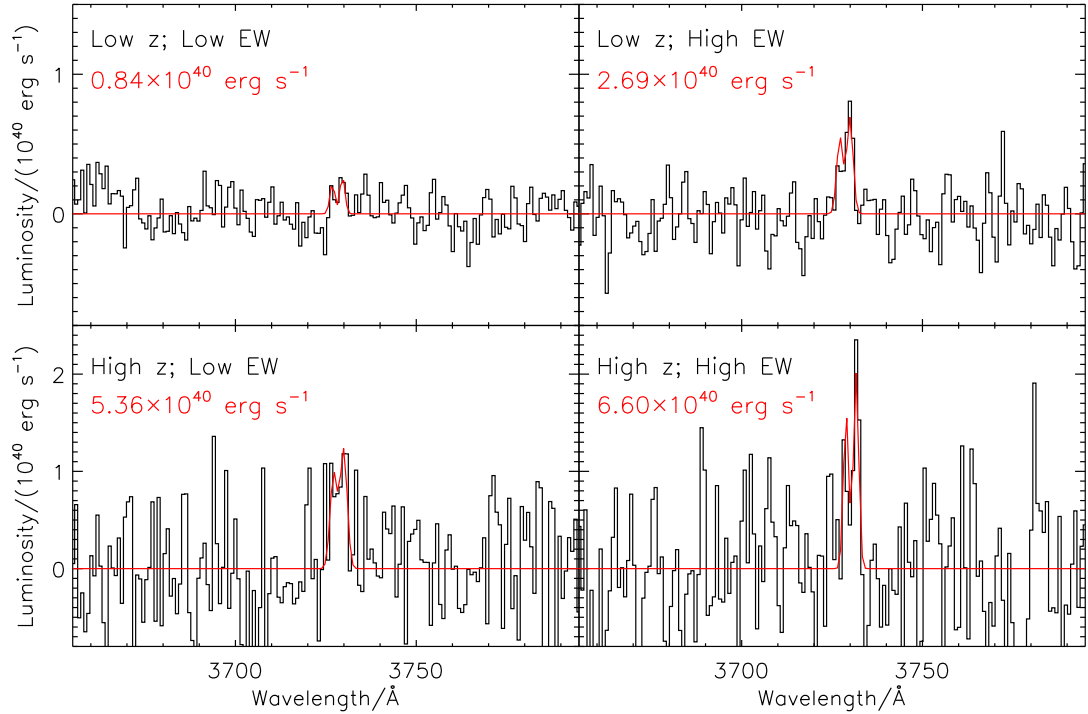}
  \end{minipage}
  \caption{Same as Fig. \ref{sfr_ca_mg} for the four subsamples of
  \caii\ absorbers split by redshift and $W_{\lambda3935}$.  No line
  is formally detected in the top left plot (SNR$<3$). Note the change in
  y-axis scale between the upper and lower plots.}
  \label{sfr_ca}
\end{figure*}

\subsection{\oii\,$\lambda\lambda${\rm 3727,3730} line fitting}

In each composite spectrum the \oii\,$\lambda\lambda$3727,3730 doublet was
fitted with a double Gaussian with line ratio fixed at 4:5\footnote{The 4:5 line
ratio is that predicted for $T$=10\,000\,K and electron density
$N_e=200\,$cm$^{-3}$ \citep{1989agna.book.....O} and it provides a good fit to the
composite \oii\,$\lambda\lambda$3727,3730 doublet.  The line strength measures
are not sensitive to reasonable variations in the adopted line ratio.}, using a
non-linear least squares fitting routine, and line luminosities measured from
the fitted Gaussian parameters.  
Whether or not a feature was detected was determined via its SNR 
\citep[][for a full derivation]{1985MNRAS.213..971H,2004AJ....127.1860B}:

\begin{equation} \label{eq:snr}
SNR =
\frac{\sum_i f_i^r u_i/\sigma_i^2}{\sqrt{\sum_i u_i^2/\sigma_i^2}} 
\end{equation}
where $f^r$ is the composite spectrum and $\sigma$ is the flux error,
$u$ is a Gaussian line profile of position and width given by the
fitted line parameters. 
A positive detection
was defined to have a SNR$>3.0$.  Where a line was
undetected, upper limits were placed for a single Gaussian line of 
$\sigma$ twice the pixel resolution and detection limit of SNR$=3.0$.

Table \ref{tab_sfr_ca} presents the measured \oii\ line luminosities and errors
for each composite.  From these values we make a few immediate observations.
Firstly, there is a significant difference between the samples split by
redshift, particularly for the \caii\ absorbers, which may be caused, in part at
least, by aperture bias --- less galaxy light falling within the fibre radius at
low redshifts.  Secondly, there appears to be a trend of increasing SFR with
increasing $W_{\lambda3935}$ in the low-redshift \caii\ subsample, but no trend
is detected at high redshift, although the errors are large.  Again, if stronger
\caii\ absorbers were more likely to be found at smaller impact parameters from
the galaxy centre, the difference could be a manifestation of aperture bias.
Due to the large errors and evidently weak, or non-existent, trends with
equivalent width, the remainder of this paper will focus on the samples split by
redshift only.

\begin{table*}
  \begin{minipage}{16.5cm}

  \caption{\label{tab_sfr_ca} \small Sample sizes, reddening, measured
  \oii\ line luminosities, inferred SFRs before and after correction
  for dust, and dust obscuration bias effecting the observed
  population, for each \caii\ composite spectrum and the two 
  \mgii-selected DLA candidate composite spectra.}

\vspace{0.2cm}

  \begin{tabular}{cccccccc} \hline\hline
  Sample & $N_{abs}$ & \EBV\ (mag) & $L$\oii\ ($10^{40}$\,erg\,s$^{-1}$) & SNR & SFR ($\Msolyr$) & SFR$^a$ ($\Msolyr$) & Dust bias$^b$ (\EBV$_{max}$) \\ \hline
  
Low $z$  & 215 & $0.036\pm0.005$ & $1.4\pm0.2$ & 5.9 & $0.09\pm0.01$ &
  $0.11\pm0.02$ & 0.24\, (0.2)\\
  
High $z$ & 130 & $0.049\pm0.004$ & $5.8\pm0.6$ & 5.8 & $0.38\pm0.04$ &
  $0.48\pm0.05$ & 0.45\, (0.25) \\
  
Low $z$, Low EW   & 153 & $0.029\pm0.006$ & $<0.9$ & 3.0 &
  $<0.06$ & $<0.07$ & 0.20\, (0.2) \\
 
Low $z$, High EW  &  62 & $0.051\pm0.009$ & $2.7\pm0.4$ & 6.6 &
  $0.18\pm0.03$ & $0.22\pm0.03$ & 0.33\, (0.2)\\
  
High $z$, Low EW  &  74 & $0.036\pm0.006$ & $5.4\pm1.3$ & 4.3 &
  $0.35\pm0.09$ & $0.41\pm0.10$ & 0.34\, (0.25) \\
  
High $z$, High EW &  56 & $0.066\pm0.007$ & $6.6\pm1.2$ & 5.0 &
  $0.43\pm0.08$ & $0.58\pm0.10$ & 0.56\, (0.25) \\

Low $z$, \mgii  & 1291 & --- & $1.7\pm0.06$ & 15.2 & $0.11\pm0.004$ &
  --- & --- \\

High $z$, \mgii & 2170 & --- & $2.1\pm0.3$ & 7.0 & $0.14\pm0.02$ &
  --- & --- \\
  \hline
  \end{tabular}\\

\vspace{0.1cm}
  $^a$ Corrected for dust attenuation at \oii, based on \EBV\ derived
  from reddening of the background quasar.\\
  $^b$ Fraction of the total population of \caii\ absorbers lost due to dust obscuration of
  the background quasar sample. This calculation is only performed for
  absorbers with \EBV\ less than the value given in brackets.
\end{minipage}
\end{table*}

\section{Star formation rates}\label{sfr}

We convert the \oii\ line luminosities into instantaneous SFRs, listed
in column 6 of Table \ref{tab_sfr_ca}, using
the relation of \citet{2004AJ....127.2002K}:
\begin{equation}
{\rm SFR([O\,{\sc II}]}(\Msolyr) = 6.58\pm1.65\times10^{-42} {\it L}({\rm [O\,{\sc
      II}]})
\end{equation}
where \oii\ luminosity is measured in ${\rm ergs\,s^{-1}}$. The errors
on our measured line fluxes and the uncertainty associated with the
SFR conversion have been propagated in the usual way.  In the
following subsections we discuss three correction factors which must
be applied to these values in order to estimate the true average SFR
of absorber host galaxies. The first two are caused by dust within the
ISM of the absorber host galaxy, the third by the finite aperture of
the SDSS fibres.

\subsection{Dust corrections to \caii\ absorbers}

While the dust content of the \mgii-selected DLA candidates is
negligible (WHP06), it is necessary to correct our measured SFRs for
the effect of dust in the \caii\ host galaxies.  This correction is
twofold.  Firstly, the \oii\ photons are absorbed by dust grains, thus
reducing the line flux emitted from the galaxy.  Secondly, the dust
obscures the background quasars, thus reducing the total redshift path
available to find \caii\ absorption line systems in the survey over
that naively expected.  To estimate the magnitude of both effects we
require the extinction as a function of wavelength caused by the dust.
This can be achieved from the measured reddening in each subset, by
assuming the form of the extinction curve and the total-to-selective
extinction ratio ($R_V \equiv A_V / \EBV$). In what follows we assume
the dust content as measured from the absorption line-of-sight is
applicable to that obscuring the nebular emission line region. In
Section \ref{sec:dust} we discuss the potential problems with this
assumption given the likely complicated geometry of the dust
distribution within galaxies.

For a full description of our method to measure dust reddening in the
absorbers see WHP06; a brief description is given here.  High SNR
quasar template spectra are created by combining the spectra of all
quasars in our input sample which do not contain known \mgii\ systems,
a total of 22\,727 quasar spectra with $0.35<\zqso<3.15$.  The
template spectra are created in redshift bins of $\Delta \zqso=0.1$.
Each quasar spectrum containing an intervening absorption line system
is then divided by a suitably normalised high SNR quasar template
spectrum.  Each residual spectrum is then moved to the rest frame of
the absorber and all residual spectra in the subsample are combined
using an arithmetic mean.  The resulting residual spectrum is fitted
with the Large Magellanic Cloud (LMC) reddening curve\footnote{WHP06
found the LMC reddening curve provided the best fit to the combined
quasar residual spectra, although the derived values of \EBV\ are not
sensitive to the exact choice of extinction curve.}  of
\citet{1992ApJ...395..130P} and the resulting $\EBV\equiv A_B-A_V$
values are given in the third column of Table~\ref{tab_sfr_ca}.

The formal errors on this fit are small and the dominant source of
error arises from the intrinsic variation in quasar spectral energy
distributions (SEDs).  This error is estimated via 10\,000 Monte Carlo
simulations drawing a random subset of quasar spectra equal in size to
the subset of absorbers.  A random absorber redshift is assigned to
each quasar spectrum and the above method repeated to estimate the
reddening of this simulated sample of absorbers.  Repeating 10\,000
times, the 68th percentiles of the final reddening distribution are
taken to provide the errors quoted in the third column of Table
\ref{tab_sfr_ca}.  Again, further details on this error calculation
are provided in WHP06.

\subsubsection{{\rm \oii}\ line attenuation}

The attenuation of light due to dust at the wavelength of \oii\ is
calculated from the measured \EBV\ values in the absorption line
systems, and the assumed extinction curve.  The resulting dust
attenuation corrected \oii\ SFRs are given in column 7 of Table
\ref{tab_sfr_ca}.  Again, the errors have been propagated in the usual
way. In the calculation we have assumed that the photons from the star
forming regions are passing through the same quantity and type of ISM
as the light from the background quasar, an assumption that will not
be correct in detail.  However, without further information on the
nature of the relation between the absorbers and host galaxies it is
difficult to improve on this assumption.  The effect on our results of
possible enhanced extinction in the nebular emission line regions is
discussed further in Section \ref{sec:dust}.

\subsubsection{Dust obscuration of the background quasars}\label{sec_bias}

The second effect of dust in the absorber is to obscure the background
quasars, meaning that fewer \caii\ absorption line systems are found
in a magnitude limited quasar sample than if the absorbers did not
contain dust.  The method of correction for this ``dust obscuration
bias'' is discussed in detail in WHP06, but given the importance of
the correction we include a revised description of the procedure in
Appendix \ref{sec:rpcorr}.

The fraction, ($b$), of the total \caii\ absorber population lost due
to dust obscuration bias is given in the final column of Table
\ref{tab_sfr_ca} for each \caii\ sample.  To obtain the true number of
\caii\ absorbers which would have been observed had they contained no
dust, the observed number is divided by $(1-b)$.  The redshift, as
well as the dust content, of the absorbers has a large effect on the
extent of the bias due to the shape of the dust extinction curves ---
the strongest attenuation is at small wavelengths; higher redshift
absorbers therefore cause greater extinction to the background quasar.
The smallest correction factor applies to the least dusty absorbers at
the lowest redshift.  Conversely for the absorbers with
$W_{\lambda3935}>0.68$ and $0.8 \le \zabs < 1.3$ we are in fact only
observing 44\% of the total number of \caii\ absorbers with
$\EBV<0.25$.  As argued in WHP06, these correction factors are very
well determined. However, as discussed in Appendix \ref{sec:rpcorr},
there is an important caveat; we are unable to correct reliably for
absorbers with an \EBV$>0.2$ and $>0.25$ for the low- and
high-redshift samples respectively because our observations have
little or no sensitivity to absorbers with such values of \EBV.


\subsection{Non-DLA contamination of \mgii-selected DLA sample}\label{noncontam}
Bias due both to \oii\ line attenuation and the obscuration of
background quasars is minimal for the \mgii-selected DLA candidate
sample given the very low dust content (see WHP06) and we have not
applied any corrections to the measured line fluxes.  One
uncertainty in the implied SFR of DLAs from our analysis lies in the
contribution made by the 50\% non-DLA contaminants to this sample.  An
estimate of the SFR of these contaminants can be calculated by
measuring the SFR found in \mgii\ absorbers which fall out of the
predicted DLA \feii/\mgii\ EW ratio range, but still with
$W_{\lambda2796}>0.6$\AA.  In these absorbers we measure SFRs of $31
(57)\%$ those found in the low- (high-) redshift \mgii-selected DLA
candidate sample, producing corrections to the SFRs quoted in the
final two rows of Table \ref{tab_sfr_ca} of a factor of 1.69(1.43).
We note that this method of estimating the correction factors is only
approximate and the associated uncertainties are discussed further in Section
\ref{sec:errors}. 


\subsection{Aperture Corrections}\label{sec_aper}

\begin{figure}
  \begin{minipage}{\textwidth}
    \includegraphics[scale=0.7]{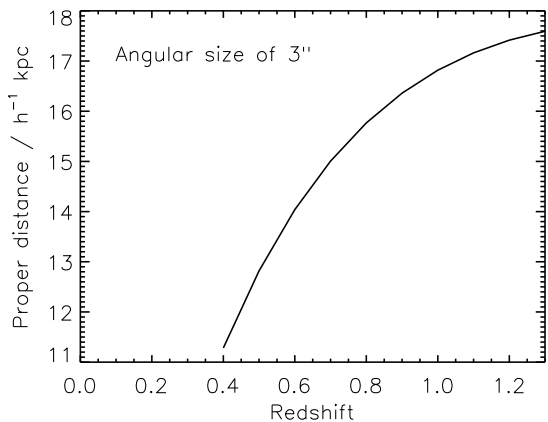}
  \end{minipage}
  \caption{Proper transverse distance covered by a 3\,arcsec diameter fibre 
   as a function of redshift, for the $\Lambda$CDM cosmology used in this
   paper with $H_0=100\,{\rm km s}^{-1} {\rm Mpc}^{-1}$.}
  \label{angdiam}
\end{figure}

The final correction factor which must be applied to our measured SFRs
in both the \caii\ and \mgii\ samples is for the finite aperture of
the SDSS fibre, which is centred on the background quasar
and not on the host galaxy of the absorber. The fibre may be covering
a fraction of empty sky, which has implications for our measures of
SFR per unit area of the absorbers; or it may be missing a fraction of the
galaxy light, with implications for our estimate of the SFR per
absorber; or both.

For ease of reference, Fig.  \ref{angdiam} depicts the proper
transverse distance covered by a 3\,arcsec diameter fibre as a function of
redshift.  Atmospheric seeing will cause a blurring of the
discontinuity at the edge of the fibre, but in practice star formation
occurring within $\sim 17/2 = 8.5\,h^{-1}$kpc from the line-of-sight
to the quasars will contribute to the observed SFR for our highest
redshift absorbers.  At the lowest redshifts in our sample this is
reduced to $\sim 14/2=7\,h^{-1}$kpc.  The fibres cover a relatively
large area, comparable to the size of the star forming extent of a
galaxy at these redshifts \citep{2004ApJ...600L.107F}, but a smaller
area than the likely extent of, for example, \mgii\ or \hi\
halos/disks surrounding the galaxies.  To calculate many quantities of
interest, such as SFR density per unit volume as a function of
redshift, $\dot{\rho}^*(z)$, information on the size of the absorbers
is required and this aspect of the discussion is deferred to Section
\ref{abssfrprop}.


\subsection{Summary of observational results}\label{obs_sum} 

Table \ref{tab_sfr_ca} summarises the principal observational result of this
paper relating to the integrated SFR for the absorbers, i.e.  the determination
of the observed average SFR for the \caii\ absorbers and the \mgii-selected DLA
candidates.  In the remainder of this section we look at the SFR per unit area
of the absorber host galaxies, for which a hard lower limit can be obtained by
dividing the observed average SFR per absorber by the area of the spectroscopic
aperture at the absorber redshift.  The lower limit arises due to the
uncertainty as to the fraction of the fibre which falls on blank sky.

\subsubsection{\mgii-selected DLAs}
By dividing the SFRs given in the final two rows of Table
\ref{tab_sfr_ca} by the proper area of the fibre at the central
redshift of the samples, and correcting for the 50\% of non-DLAs in
the sample as discussed above, we obtain limits on the SFR per unit
area of DLAs.  The limits are $\gsim 12\pm0.4\times10^{-4}$ and $\gsim 9\pm1 \times 10^{-4}\,{\it
h}^2\,\Msolyr\,\perkpcsq$ for the low-$z$ ($\langle z\rangle\approx0.6$)
and high-$z$ ($\langle z\rangle\approx1.0$) samples.

To put these limits into context, we compare with the Kennicutt
parameterisation of the Schmidt law (Schmidt 1959; Kennicutt 1998a; 1998b)
\nocite{1959ApJ...129..243S,1998ARA&A..36..189K,1998ApJ...498..541K}
which relates the surface densities of neutral gas and SFR in
galaxies: 
\begin{equation}\label{eq:schmidt}
\sum _{\rm SFR} = 2.5\times10^{-4} \times
(\frac{N({\rm H I})}{1.25\times10^{20}\,\percmsq})^{1.4}
\end{equation}
above a critical \hi\ column density, $N_c = 5 \times
10^{20}\,\percmsq$, where $\sum_{\rm SFR}$ is in units of
$\Msolyr\,\perkpcsq$.  A distribution function for the \hi\ column
densities of the absorbers is thus required to predict their SFR per
unit area.  In fact, the combination of the critical threshold value
$N_c$ for the onset of star formation, combined with the non-linear
dependence of the SFR on column density, means that simply evaluating
Equation \ref{eq:schmidt} for $\langle$\nhi$\rangle$ produces an
answer accurate to better than 10\%.

Taking the distribution of \hi\ column densities for DLAs at
$1.7<z<2.2$ of \citet{2005ApJ...635..123P} results in a mean column
density of $6.6 \times 10^{20}\,\percmsq$.  The mean is insensitive
both to the exact form of the acceptable functional fits given in
\citet{2005ApJ...635..123P} and the upper cutoff in \hi\ column
density between $10^{22}$ and $10^{23}\,\percmsq$.  The latter
behaviour is due to the steep decline in the number of systems at very
high column densities.  Unfortunately, at $z<2$, the redshift range
applicable to our systems, significant uncertainty in the predicted
SFR per unit area for the absorber populations comes about because of
the poorly determined distribution of
\nhi. \citet{2006ApJ...636..610R} find that the mean \nhi\ is somewhat
higher than assumed above, $\langle$\nhi$\rangle = 1.27\pm 0.36
\times10^{21}\,\percmsq$ for DLAs with $0.11<z<0.9$, and $1.07\pm 0.23
\times10^{21}\,\percmsq$ for DLAs with $0.9<z<1.65$.

The \citet{2006ApJ...636..610R} mean column density
$\langle$\nhi$\rangle$ corresponds to a SFR per unit area of
$13.1\pm5.2 \times 10^{-3}$ and $10.3\pm3.1 \times 10^{-3}\,{\it
h}^2\,\Msolyr\,\perkpcsq$ according to Equation \ref{eq:schmidt}, for
their low and high redshift bins respectively.  These values are
factors of $11\pm4$ and $11\pm3$ higher than the lower limits we
deduced for our low- and high-$z$ samples (the errors here are
dominated by the uncertainty in $\langle$\nhi$\rangle$).  For
comparison, the high redshift $\langle$\nhi$\rangle$ value of
\citet{2005ApJ...635..123P} results in a predicted SFR per unit area
for DLAs of $5.2 \times 10^{-3}\,{\it h}^2\,\Msolyr\,\perkpcsq$.
This is still 4(6) times greater than the derived lower limit in the low-
(high-) $z$ \mgii-selected DLAs.

Does the mean \hi\ column density of DLAs apply to our \mgii-selected
sample?  These absorbers were selected to maximise the
DLA fraction they contain and allowance was made to correct the
observed SFR for contamination by non-DLAs (Section \ref{noncontam}).
The results of \citet{2006ApJ...636..610R} are based on follow-up
\nhi\ observations of a sample of \mgii-selected systems very similar
to that used here and it would be very surprising if the properties of our
\mgii\ absorber sample were significantly different from those studied by
\citet{2006ApJ...636..610R}. 

\subsubsection{\caii\ absorbers}
In the case of the \caii\ absorbers, dividing the SFRs by the proper
area of the fibre at the central redshift of the samples results in
limits for the SFR per unit area of $\gsim 7 (21) \times
10^{-4}\,{\it h}^2\,\Msolyr\,\perkpcsq$ for the low- (high-) $z$ samples
respectively, after correcting the SFRs for attenuation of the \oii\
line by dust.

It is difficult to compare these results to those predicted by the
Schmidt law, due to the very few systems currently with both \caii\
and \nhi\ measures, particularly for the high $W_{\lambda3935}$
systems.  While the \caii\ absorbers are expected to have \nhi\ values
above the nominal limit for DLAs, the comparison to predicted SFRs per
unit area of DLAs should be treated with caution.

In the following section, we discuss the magnitude of corrections
applicable to these values caused by portions of the fibre aperture
covering blank sky, and convert the measured lower limits into
absolute SFR per unit area values for the absorbers. Here we simply
note that the corrections are expected to be small, particularly in
the case of the \mgii-selected DLAs.


\section{Absorber star formation properties}\label{abssfrprop}

In this section we use results of imaging surveys of DLAs to estimate
their average extent and thus assess the magnitude of corrections
applicable to our observed SFRs within the fibre aperture.  The
discussion in this section is predicated on the assumption that the
SFR per unit area is determined entirely by the Schmidt-law via the
\hi\ column density, as specified in Equation \ref{eq:schmidt}. If
DLAs at $z \la 1$ are caused by large gas disks surrounding a central
galaxy, the absolute SFR per unit area observed through the SDSS fibre
aperture can be predicted from the mean \nhi\ of the population and,
furthermore, we can estimate the contribution of the host galaxies to
the star formation rate density ($\sfrd$) of the universe.  Discussion
of scenarios in which the relative location of star formation and
absorption within a galaxy/halo system, and/or failure of the Schmidt
law, impact greatly on our results follows in Section
\ref{discussion}.

\subsection{Physical size of absorbers}\label{sec:size}

In the case of DLAs, imaging of host galaxies at $z\lsim1$ suggests
they may occur in halos or disks that extend out as far as
$\sim25h^{-1}$ kpc from the host galaxy (e.g.  Chen \& Lanzetta 2003;
Rao et~al.  2003; see Table 2 of Zwaan et~al.  2005 for a recent
compilation)\nocite{2003ApJ...597..706C, 2003ApJ...595...94R,
2005MNRAS.364.1467Z}.  The mean observed impact parameter between
DLAs and absorber host galaxies is $\sim 9 h^{-1}$\,kpc \citep[Table 2
of][]{2005MNRAS.364.1467Z}.  Assuming a simple circular geometry for
the absorbers this corresponds to a typical absorber radius\footnote{The mean
observed impact parameter is about two-thirds of the maximum radius,
assuming the absorbers present simple circular cross-sections.}  of
$\simeq 13 h^{-1}$\,kpc, assuming covering factors of unity.
Comparatively, from direct \hi\ observations of local galaxies,
\citet{2005MNRAS.364.1467Z} predict a very similar mean impact parameter of
$\sim7 h^{-1}$\,kpc.  

We therefore adopt $13 h^{-1}$\,kpc as the predicted extent from the
centre of the host galaxy of absorbing clouds which give rise to DLAs.
Adopting the measured value of $dP/dz = 0.079 \pm 0.019$
\citep{2006ApJ...636..610R} for DLA absorbers at low redshifts, the
estimated size leads to absorber space densities of $0.027
(0.022)\,h^3\,{\rm Mpc}^{-3}$ for our low- and (high-) redshift
samples (Equation \ref{eq:dpdz}).

We turn now to the \caii\ absorbers.  Without any direct observations
of host galaxies, or information on how far \caii\ extends into the
gaseous halos surrounding galaxies, we must make some assumption about
their size.  In WHP06 strong arguments are given that \caii\ absorbers
have \nhi\ values above the nominal limit for DLAs and, as such, make
up a subsample of DLAs.  Their high dust and metal contents suggest a
scenario in which \caii\ absorbers exist closer to the central galaxy,
possibly with a significant fraction lying within the star forming
region of the galaxy.  The values of
$dP/dz\,(W_{\lambda3795}>0.5\,{\rm \AA})$ for the \caii\ absorbers,
correcting for those missed from the sample due to dust obscuration
bias, are 0.019 and 0.025 for our low- and high-redshift samples
(WHP06 and see Section \ref{sec:cross}).  Taking the same space
densities as for the DLA absorbers and simply scaling the
cross-sections according to the difference in the observed $dP/dz$ for
the two absorber populations gives radii for \caii\ absorbers of $0.49
(0.56)$ times those of DLAs, or $6.4 (7.3) h^{-1}$kpc for the low-
(high-) $z$ samples\footnote{Despite the different method used to
estimate the halo sizes of \caii\ absorbers, the resulting size
remains very similar to that suggested in WHP06.}.  We note that the
difference in size for the two redshift intervals is not significant
given the combined errors on number density for both the DLAs and
\caii\ absorbers.

\subsection{SFR per unit area} \label{sec:sfrarea}

Given the physical size of the absorbers we can calculate their actual
SFR per unit area, within our assumed geometrical model, for which
lower limits were calculated in Section \ref{obs_sum}.

Beginning with the \mgii-selected DLA candidates, placing
$6-9\,h^{-1}$kpc radius circles (i.e.  the SDSS spectroscopic fibres,
see Section \ref{sec_aper}) at random across a second circle of radius
$13h^{-1}$kpc (i.e.  the gaseous halos or disks), such that the impact
parameter between the centre of the circles never exceeds the radius
of the second circle, and summing the fraction of area falling outside
the absorber boundary (i.e.  on blank sky) leads to aperture
corrections of 1.16 and 1.21 for low- and high-$z$ samples
respectively.  Thus, the SFR per unit area of \mgii-selected DLA
candidates derived in Section \ref{obs_sum} can be increased only to
$14\pm0.5(11\pm1.6) \times 10^{-4}\,{\it h}^2\,\Msolyr\,\perkpcsq$ and a large
discrepancy between the conventional Schmidt-relation and our
observations remains.

Adopting the estimates for the size of the \caii\ absorbers leads to
an aperture correction of 1.59 (1.69) for the low- (high) $z$ samples.
These corrections lead to a SFR per unit area of $11(36) \times
10^{-4}\,{\it h}^2\,\Msolyr\,\perkpcsq$, i.e.  very similar to the
\mgii-selected DLA candidates at low redshift, but four times higher
in the high redshift sample.  {\it If} the mean \hi\ column of \caii\
absorbers is equal to that of DLAs with $1.7<z<2.2$, the high redshift
\caii\ absorber host galaxies have SFRs per unit area closer to that
predicted by the Schmidt law, while at low redshifts an order of
magnitude discrepancy remains.

\subsection{The global SFR density}

The global SFR density of the universe has now been measured by many
surveys out to high redshift. The general consensus is that, as we look
back in time, the SFR density increases by about one order of magnitude
from $z = 0$ to 1, and remains approximately constant at higher
redshifts up to $z \sim 5$ \citep{2006NewAR..50..152B}.
\citet{2004ApJ...615..209H} has provided a power-law fit
to the data at $z<1$:

\begin{equation}\label{eq:sfr}
\log \sfrd = 3.29 \log (1+z)-1.8
\end{equation}
where $\sfrd$ is in units of $\Msolyr\,{\rm Mpc^{-3}}$ and
$H_0=70\,\kms\,{\rm Mpc}^{-1}$.  In our low- (high-) $z$ bins the
average SFR density of the universe is thus $\log \sfrd \simeq-1.13$
($-0.81$) or, with $\sfrd$ in units of ${\it h}^3\,\Msolyr\,{\rm
Mpc^{-3}}$, $\log \sfrd \simeq-0.66(-0.35)$. From now on all $\sfrd$
values will be quoted in the latter units.

Adopting either a physical size or a space density for the \caii\ and
\mgii-selected DLA absorbers, and assuming the star formation is
uniformly distributed (on scales similar to the fibre size) within the
high column density gas, as required by the Schmidt law, we can
estimate the contribution to the global SFR density of the universe
from the host galaxies of each class of absorbers.

\subsubsection{Cross sectional area and comoving space density of the absorbers}\label{sec:cross}

\begin{figure*}
  \begin{minipage}{\textwidth}
    \includegraphics[scale=0.65]{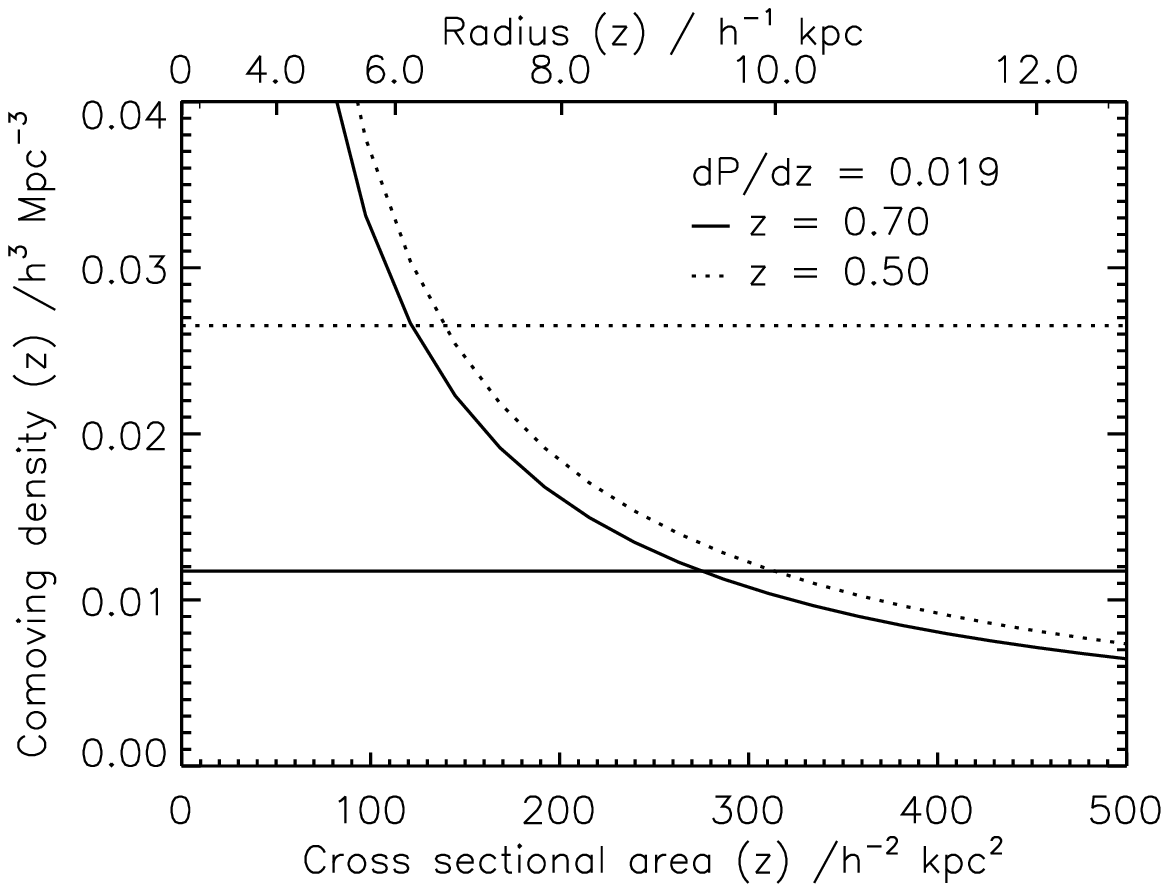}
\hspace{1cm}
    \includegraphics[scale=0.65]{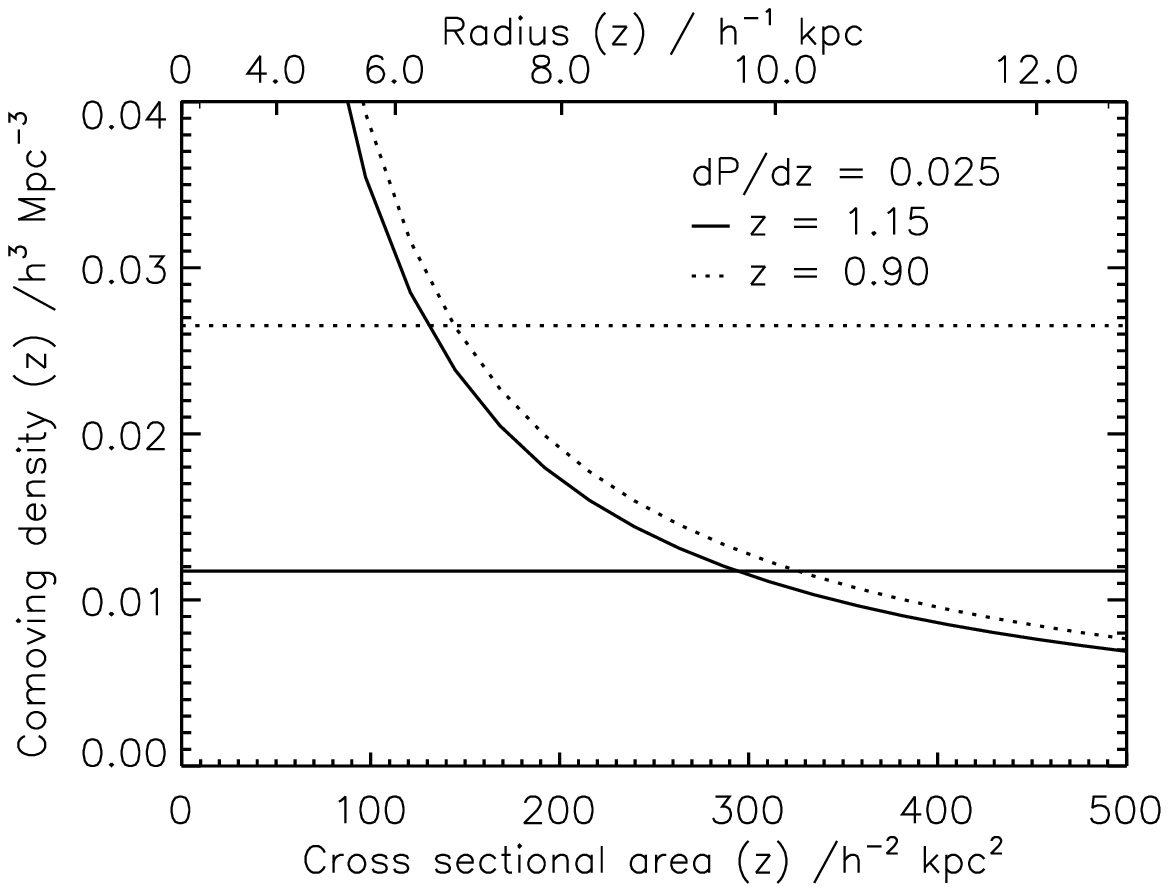}
  \end{minipage}
  \caption{Combined constraints on cross sectional area and comoving
  number density, given the measured $dP/dz$ of the \caii\ absorption
  line systems corrected for those missed from the sample due to dust
  obscuration bias. {\it Left}: low redshift sample; {\it right}: high
  redshift sample. In each plot, the constraints are shown for two
  different redshifts within the redshift bin. The horizontal lines
  are comoving number densities of galaxies, at $z\sim 0$, with $M>M^*+1$
  and $M>M^*+2$, where $M^*$ is the 2dFGRS characteristic
  magnitude (see text).}
  \label{fig_number}
\end{figure*}

The observed redshift number density ($dP/dz$) of absorbers provides a
joint constraint on their size and space density.  For absorbers with
cross-sectional area, $S$, and comoving number density, $n(z)$, their
number per unit redshift is given by
\citep{Hogg_distances99,1993ppc..book.....P}:

\begin{equation}\label{eq:dpdz}
\frac{dP}{dz} = S\ n(z)\ \frac{c}{H_0} 
\frac{(1+z)^2}{\sqrt{\Omega_M(1+z)^3+\Omega_\Lambda}}
\end{equation}

The $dP/dz$ of the whole \caii\ absorber population, i.e.  correcting for dust
obscuration bias (Section \ref{sec_bias}), is obtained by multiplying the
observed $dP/dz$ by the fraction lost due to dust bias (final column of Table
\ref{tab_sfr_ca}).  For the low- and high-redshift samples we find
$dP/dz\,(W_{\lambda3795}>0.5\,{\rm \AA})=0.019$ and $0.025$
respectively\footnote{Strictly speaking these are lower limits to the true
number densities, as we are unable to correct for dust obscuration bias above
the $\EBV_{max}$ values given in Table \ref{tab_sfr_ca}.}  (see WHP06 for a
detailed method).  In Fig.  \ref{fig_number} we plot the combined constraint on
comoving number density and cross sectional area of the dust bias corrected
\caii\ absorber samples, given their respective redshift number densities.  The
constraints for two different redshifts within each redshift interval are
plotted, although the change within a bin is not large.

To provide reference points, the $z$$\sim$0 space density of galaxies
with $B$-band magnitudes, $M > M^*+1.0$ and $M >M^* +2.0$ are
indicated by horizontal lines.  $M^*= -19.79+5\log_{10}(h)$ and
$\phi^* = 0.00159 h^3\,{\rm Mpc}^{-3}$, the characteristic magnitude
(on the Vega-system) and normalisation of the luminosity function, are
taken from the rest-frame $B$-band Schechter function parameterisation
of the 2dF Galaxy Redshift Survey
\citep[2dFGRS,][]{2002MNRAS.333..133M}.

\subsubsection{Contribution of absorber host galaxies to
$\sfrd$}\label{rhodot} 

Starting with the \caii\ absorbers, within the errors, the space
density estimates of $0.027(0.022)\,h^3\,{\rm Mpc}^{-3}$ (Section
\ref{sec:size}), for the low- (high-) $z$ samples, are consistent with
zero evolution and are similar to the space density of galaxies
brighter than $M^*_{\rm 2dF}+2$.

Given the small impact parameters expected for the \caii\ absorbers
within our assumed model for the relative configuration of star
formation and absorbing gas, the central host galaxy is expected to
fall within the fibre aperture essentially all the time (Fig.
\ref{angdiam}).  The fibre aperture correction factor (that quantifies
the host galaxy emission line flux lying outside the fibre diameter)
is
1.32(1.25) for the low- (high-) $z$ samples.  The volume averaged
SFR density of the absorber host galaxies, obtained from the product
of the SFR, the space density and the aperture corrections, is then
$\log\sfrd\simeq -2.43(-1.89)$.
Comparing to Equation \ref{eq:sfr} these values represent only
2(3)\% of the total SFR density in the universe at the same
redshifts.

Turning now to the \mgii-selected DLA absorbers, $dP/dz$ is a factor
three to four larger and, unless the DLAs possess a much higher space
density than that of the \caii\ absorbers, allowing them to be
considerably smaller, their physical extents are such that a
significant aperture correction is necessary in order to allow for the
absorber gas that falls outside the 3\,arcsec SDSS fibre aperture.
Adopting the absorber size estimate of $13 h^{-1}$\,kpc, and hence
space density, from Section \ref{sec:size} leads to aperture
corrections of 3.9(2.9) for the low- (high-) $z$ samples respectively.
Taking the measured SFR for the \mgii-selected DLA candidate sample
then leads to values of $\log\sfrd\simeq -1.72(-1.91)$.  These values correspond to
$\sim$9(3)\% of the SFR density in the universe at $<$$z$$>\simeq$0.6
and $<$$z$$>\simeq$1.0 respectively.

It is not possible to explain the small contributions to the global
SFR density by simply appealing to an underestimate of the size of the
absorbers in Section \ref{sec:size}. Once absorber size is comparable
to or larger than the fibre-size (as in all examples here), and {\it
assuming the Schmidt law holds}, changes in the assumed absorber size
make no significant difference to the value of $\sfrd$. This follows
from $\sfrd$ being proportional to the product of the comoving number
density and aperture correction:
\begin{eqnarray}\label{eq:6}
\sfrd & \propto& \langle {\rm SFR} \rangle \times n(z) \times A\\\nonumber
&\propto & \langle {\rm SFR} \rangle \times \frac{dP/dz}{S} \times A
\end{eqnarray}
where $\langle$SFR$\rangle$ is the SFR per absorber, $A$ is the
aperture correction factor, $n(z)$ the comoving number density, $S$
the area of the absorber and $dP/dz$ is fixed by observation. In the
limit that the size of the absorber (and star formation region) is
greater than the fibre aperture, as $S$ increases the aperture
correction increases proportionately and $\sfrd$ remains constant.

Our results for the $\sfrd$ of \mgii-selected DLA candidates are
directly comparable to those of \citet{2005ApJ...630..108H} who
concluded that DLAs are responsible for as much as 80\% of the SFR
density of the universe at similar redshifts to those considered here.
The dramatic discrepancy between our study and theirs arises directly
from the assumption by Hopkins et~al.  that the local Schmidt relation
holds for DLA absorbers. Assuming the mean \nhi\ for our
\mgii-selected DLAs from \citet{2006ApJ...636..610R}, as appropriate
for a direct comparison between the results, we observe a SFR per unit
area of $\la$10\% of that expected were the Schmidt law to apply.

\subsection{Summary of uncertainties}\label{sec:errors}

For each of our absorber samples potential sources of uncertainty in
our analysis have been made clear throughout the paper. The final SFR
measurements presented in Table 1 include associated line measurement
errors, the error on the conversion from \oii\ luminosity to SFR and,
in the case of the \caii\ absorbers, the error on the dust attenuation
correction estimated from the reddening of the background quasar
continuum. In this subsection we combine these errors with additional
uncertainties in our calculations to produce an estimate of the
reliability of our key results.

\subsubsection{Star formation rates from \oii\ emission}

\citet{2004AJ....127.2002K} present a comprehensive analysis of the
dependence of the \oii-derived SFR in systems with an extended range
of physical properties, including objects with SFRs significantly
lower than those we measure for the absorber sample. Given the known
low metallicity of DLA gas, the adopted \citet{2004AJ....127.2002K}
relation applies for systems with a mean metallicity that may be
higher than that of the absorber populations.  Applying a correction
for metallicity would {\it decrease} the inferred SFR for a given
\oii-luminosity, increasing the disagreement between our observations
and predictions based on the application of the Schmidt law. As very
few absorption systems have measured emission line abundances, and the
link between absorption and emission line metallicities therefore
remains unclear, it is not possible to assess this effect further at
this stage.
  
\subsubsection{Dust attenuation of the \oii\ emission}\label{sec:dust}

In calculating the SFR of the \mgii-selected DLAs we have not made any
correction for dust attenuation of the \oii\ line, because of the very
low reddening detected in \mgii\ absorbers by WHP06. Specifically, for
an equivalently defined sample we found an \EBV\ of $0.008\pm0.0016$
which results in a flux correction of $<$4\% at 3729\AA, insignificant
compared to the line measurement errors.

Unlike the \mgii\ absorbers, the \caii\ absorbers do contain
significant quantities of dust. However, as argued in WHP06, the
quantity of dust, associated errors and resulting obscuration bias are
well determined. The \EBV\ values obtained are not dissimilar to those
expected from sightlines passing through the disks of spiral galaxies.

However, it is possible, and perhaps likely, that nebular emission
line regions suffer significantly greater dust extinction than that
experienced by the majority of the interstellar
medium. \citet{2001ApJ...548..681B} find a median \EBV\ of $\sim0.4$
for nebular emission regions of nearby star forming galaxies, although
the mean SFR of their sample is considerably larger than for the
systems studied in this paper.

With the currently available data it is only possible to obtain a
rough estimate of the true attenuation in the nebular emission regions
we observe. By combining all 1559 \mgii\ absorbers with $\zabs<0.85$,
in which H$\beta$ is potentially visible, we find an \oii\ to H$\beta$
ratio of 1.9. Given the theoretical balmer decrement
\citep{1989agna.book.....O}, relation between H$\alpha$ line strength
and SFR \citep{1998ARA&A..36..189K}, and between \oii\ line strength and SFR
\citep{2004AJ....127.2002K}, we predict a ratio of 3.44. This suggests
dust attenuation may cause us to be underestimating the SFRs by up to a
factor of two, however, we emphasise that an accurate analysis must
await further observations.

\subsubsection{\mgii-selected DLAs}

Firstly, we look at the result with perhaps the farthest reaching
implications, that the SFR per unit area in DLA gas is observed to be
a magnitude lower than that expected from the Schmidt law. We consider
in turn the effects of the adopted mean \nhi, DLA fraction, absorber
size, contamination by non-DLA absorbers to the measured SFR and dust.

As stated in Section \ref{obs_sum} it would be very surprising were
the mean \nhi\ and contamination fraction of our \mgii-selected DLA
sample to turn out to be very different from those found by
\citet{2006ApJ...636..610R} given the essentially identical criteria
used to define both absorber samples.
Were we to overestimate the size of the absorbers we would
underestimate the SFR per unit area. The scenario in which the
absorber size is reduced to bring our result into line with the
Schmidt law is discussed in Section 5 and found to be highly
implausible.

Turning to the contamination of the sample by non-DLA absorbers, we
concluded in Section \ref{noncontam} that non-DLAs could have SFRs of
$\sim31-57\%$ that of the DLAs. If, however, the non-DLAs contributed
zero SFR to the measured lines our SFR per absorber would increase by
only a further 20\% and 40\% for the low- and high-redshift samples
respectively.  Combining the final SFR per unit area of Section
\ref{sec:sfrarea} with the possibility that the non-DLAs in our sample
contribute zero flux to the observed lines allows a SFR per unit area
of at most $16.6\pm0.6(15\pm2)\times 10^{-4}\,{\it
h}^2\,\Msolyr\,\perkpcsq$, i.e. still a factor of eight (seven) below
the Schmidt law for the low- (high-) $z$ samples respectively.

Finally, as discussed in the previous subsection, attenuation of the
\oii\ line by dust may be our biggest cause of uncertainty, due to the
unknown distribution of dust around star forming regions compared to in
the interstellar medium. Allowing for an underestimate of the SFR by a
factor of two, and that the non-DLAs contribute zero flux, would
reduce the discrepancy between our results and the Schmidt law to a
factor of four (three).

Turning now to the SFR density measurements, assuming the Schmidt law
holds and allowing for the non-DLA contaminants to contribute zero
flux to the measured lines, we find maximum possible contributions of
DLA host galaxies to the overall SFR density of the Universe of
$10\pm3\%$ ($4\pm1\%$). Within the adopted model, an underestimate of the
absorber size does not change our results, due to the compensation by
the aperture correction (Equation \ref{eq:6}). Allowing for the
possibility that we have underestimated dust attenuation by at most a
factor of two brings this result to $20\%$ ($8\%$).

\subsubsection{\caii\ absorbers}

As with the \mgii-absorbers, an underestimate of the absorber size can
not change our results, provided the Schmidt law applies. Even given
the uncertainty in dust extinction, it is difficult to see how the
inferred contribution of \caii\ absorber host galaxy light, that falls
within the $\sim 15 h^{-1}\,$kpc diameter covered by the SDSS-fibres,
can contribute more than a few percent to the global SFR density.

\section{Discussion}\label{discussion}

Given our lack of knowledge concerning the size distribution and
geometry of \caii\ and DLA absorbers we have, in the previous section,
deliberately confined ourselves to providing illustrative calculations
of the SFR per unit area and the volume averaged SFR density based on
conventional assumptions concerning the size of the absorbers and the
relationship between $N$(\hi) and SFR.  In particular, no account has
been taken of more realistic geometries that may be involved in the
relation between the measured $N$(\hi)-column along a single
line-of-sight through an absorber and the spatial distribution of star
formation within the system giving rise to the absorber.  Further
quantitative progress must await the results of imaging investigations
capable of constraining the extent of the star forming regions in
\caii\ and DLA absorbers at $z$$\lsim$1.  However, the detection of
weak \oii\,$\lambda\lambda$3727,3730 nebular emission associated with
the \caii\ and \mgii-selected DLA candidate absorbers places strong
constraints on the combination of absorber size, space density and
star formation properties of absorption selected systems.

The lower value of $dP/dz$ and the significantly higher SFR seen in
the high redshift \caii\ absorbers compared to the \mgii-selected DLA
candidate absorbers fits in well with the view advanced in WHP06 that
the \caii\ absorbers represent chemically evolved systems with
detectable quantities of dust that may be spatially closely related to
intermediate brightness late-type galaxies.  That the combination of
absorber size and SFR produces a SFR per unit area closer to that
expected from the Schmidt law, assuming \caii\ absorbers have mean
\nhi\ similar to that found in DLAs, adds further weight to such an
interpretation.  However, the consequence of this picture is that
\caii\ absorbers then account for only a small fraction, just $\sim
3$\%, of the total SFR density in the universe at $z \simeq 0.5 - 1$.
Appealing to a missing fraction of absorbers with associated
extinction exceeding our sensitivity limit of $\EBV\simeq0.25\,$mag
can increase the fractional contribution to the $\sfrd$ but not to the
extent of explaining the factor of $\sim$30 discrepancy.  Indeed, the
fraction of ``missing'' absorbers required would be inconsistent with
established limits on the number of obscured DLA systems at high
redshift \citep{2001A&A...379..393E, 2004ApJ...615..118E}.

Although the contribution of the low- and high-redshift \caii\ samples
to the global SFR density is constant, the significant difference
between the observed SFR per unit area for the two samples is
something of a puzzle. A resolution to the difference awaits the
results of imaging studies of the \caii\ absorbers and we will report
on such observations in the near future.

The low measured SFR associated with the \mgii-selected DLA absorbers
results in new constraints on the SFR per unit area and, assuming the
absorbers are large ($\gsim 9 h^{-1}\,$kpc), on the contribution to
the volume averaged SFR density, $\sfrd$, of the host galaxy light
falling within a $\simeq 7.5 h^{-1}\,$kpc radius of the DLA absorbers.
If the absorbers are indeed large, and assuming the
\citet{2006ApJ...636..610R} mean value for \nhi, then the SFR per unit
area may be as much as an order of magnitude, and certainly a
factor of several, below that predicted
by the Schmidt law. Correspondingly, the contribution of DLAs to
$\sfrd$ (at $z$$\sim$1) of the universe is dramatically smaller than
that deduced by \citet{2005ApJ...630..108H}, who assumed that the
Schmidt law does apply.

It is possible to make our constraints on the SFR per \mgii-selected
DLA candidate consistent with the predictions of the Schmidt law, and
hence produce a contribution to $\sfrd$ as derived by
\citet{2005ApJ...630..108H}, but only by making the individual DLA
absorbers very small.  Specifically, to bring the SFR per unit area
for the \mgii-selected DLA candidate absorbers into agreement with the
Schmidt prediction and $\langle$\nhi$\rangle$ of
\citet{2006ApJ...636..610R}, the physical size of the absorbers must
be $\simeq 3(4)h^{-1}$\,kpc in radius, where the absorber size comes
about by requiring that the surface area of an absorber times the SFR
per unit area predicted by the Schmidt law, equals the integrated SFR
per absorber we observe (Table 1).  The size of the actual regions
associated with star formation (i.e.  with \nhi$ > 5 \times
10^{20}\,\percmsq$), is then only $\simeq 2(2.5)h^{-1}$\,kpc in
radius. The small physical size of the absorbers then requires that
the space density of DLA absorbers at $z$$\sim$1 is very large indeed,
$\sim 0.13 h^3 {\rm Mpc^{-3}}$. A scenario in which such small
absorbers are hosted by galaxies of similar space density is
inconsistent with the established observation that the bulk of star
formation at $z$$\sim$1 occurs in luminous galaxies
\citep{2003A&A...402...65H,2005ApJ...630..771W,2005ApJ...625...23B}.
Alternatively, gaseous galaxy halos may contain many small DLA
absorbers (i.e. have ``filling factors'' considerably less than
unity), extending far beyond the extent of the SDSS fibre. Such a
scenario is, however, inconsistent with observed DLA and associated
metal line profiles as many independent components would be visible
within the profiles \citep{2005ARA&A..43..861W}. In summary,
reducing the size of the absorbers such that Schmidt law holds, produces a
size/space-density constraint inconsistent with any current favoured
model for DLA absorbers and galaxies in general. 

Alternatively, the low-redshift Schmidt law may not hold for gas
contained in intermediate to high redshift quasar absorption line
systems. Recently, Wolfe \& Chen (2006) have deduced an upper limit to
the SFR per unit area associated with DLA absorbers at redshifts
$z=2.5-3.5$ of less than 10\% of that predicted from the Schmidt law.
The Wolfe \& Chen limit is based on the non-detection of the absorbers
in the Hubble Ultra-Deep Field and is therefore a strict {\it upper}
limit.  Wolfe \& Chen's interpretation is that in galaxies with low
dust and molecular content, as in the DLAs, the threshold \hi\ column
density required to trigger star formation is significantly higher
than in the nearby galaxies where the empirical Schmidt law was
calibrated. Both these results, and the results presented in this
paper, are inconsistent with the SFR per unit area derived from C~{\sc
ii}$^*$ by \citet{2003ApJ...593..235W}; for a full discussion of
possible reasons we refer the reader to Wolfe \& Chen (2006).

Whatever the physical explanation, it is now difficult to escape the
conclusion that only a small fraction of the star formation rate seen
directly in galaxy surveys at redshifts from $z$ = 0.5 to 3.5 is
apparently associated with the relatively large 6-9 $h^{-1}$kpc region
surrounding DLA absorbers.  Evidently, the largest contribution to the
DLA cross-section is from gas which is too diffuse to support high
rates of star formation and metal production, thus explaining the
generally low metallicities of most DLAs
\citep[e.g.][]{CJA05}. However, there are two plausible models
consistent with our results. Recent theoretical work by Johansson \&
Efstathiou (2006) favours extended distributions of DLA gas, with the
cross section dominated by small galaxies, and their predicted mean
SFR per DLA host is in very good agreement with the values measured
here. Recent observational work at both low and high redshift support
such a scenario: \citet{2006AJ....131..686C} measure luminosities of low
redshift DLA hosts of $\le0.1L*$; and \citet{2006ApJ...636...30K}
place limits on the SFR of high redshift DLA hosts of
$0.9-2.7\Msolyr$. Alternatively, Wolfe \& Chen (2006) associate low surface density
\hi\ regions, in which a reduced star formation rate efficiency
exists, with the periphery of luminous, star-forming galaxies at $z
\simeq 3$. The low redshift Schmidt law may indeed still apply in the
inner regions of these galaxies but the typical host to DLA absorber
separation must then exceed the $\simeq 7.5 h^{-1}\,$kpc radius corresponding
to the SDSS fibres.  It should be relatively straightforward to assess
the validity of these models at $z < 1$, where the putative
star-forming galaxies associated, but not coincident, with the DLAs
should be relatively easy to identify from ground-based images.  We
have begun such a study and will report on the results in the near
future.


\section*{acknowledgments}

We are grateful to Bob Carswell, Hsiao-Wen Chen, Emma Ryan-Weber and
Art Wolfe for helpful discussions and to the anonymous referee for
suggestions which improved the paper.  VW is supported by the MAGPOP
Marie Curie EU Research and Training Network.

Funding for the SDSS has been provided by the Alfred P.  Sloan
Foundation, the Participating Institutions, the National Science
Foundation, the U.S.  Department of Energy, the National Aeronautics
and Space Administration, the Japanese Monbukagakusho, the Max Planck
Society, and the Higher Education Funding Council for England.  The
SDSS Web Site is http://www.sdss.org/.

The SDSS is managed by the Astrophysical Research Consortium for the
Participating Institutions.  The Participating Institutions are the
American Museum of Natural History, Astrophysical Institute Potsdam,
University of Basel, Cambridge University, Case Western Reserve
University, University of Chicago, Drexel University, Fermilab, the
Institute for Advanced Study, the Japan Participation Group, Johns
Hopkins University, the Joint Institute for Nuclear Astrophysics, the
Kavli Institute for Particle Astrophysics and Cosmology, the Korean
Scientist Group, the Chinese Academy of Sciences (LAMOST), Los Alamos
National Laboratory, the Max-Planck-Institute for Astronomy (MPIA),
the Max-Planck-Institute for Astrophysics (MPA), New Mexico State
University, Ohio State University, University of Pittsburgh,
University of Portsmouth, Princeton University, the United States
Naval Observatory, and the University of Washington.

\bibliographystyle{mn2e}


\begin{appendix}
\section{Correcting the redshift path for the effect of dust 
associated with the absorbers}\label{sec:rpcorr}

The probability of finding an absorption line system intervening in a quasar
spectrum depends on the total redshift path between the quasar and us, the
signal-to-noise ratio (SNR) of the quasar spectrum and the equivalent width of
absorber that we are searching for.  The effect of dust in an intervening
absorption system is to reduce the SNR of the quasar spectrum, thus reducing the
effective redshift path-length that each spectrum provides for finding an
absorber of given equivalent width.  The quasar may also be entirely removed
from the sample as its magnitude falls below the survey flux limit.

To estimate the total effect of a population of absorbers with known
distributions of line equivalent width, dust content and redshift, on
the background quasar population of our survey, we simply imagine
placing each absorber in front of every one of the input quasars in
our survey. We calculate the redshift path available to find a
dust-free absorber:
\begin{equation}
(\Delta z)_{j,no-dust} = \sum_{k, m_k<19.1} \Delta X_k \times P(z_{abs,j}, 
W_j, m_k)
\end{equation}
and that available to find a dusty absorber causing $A$ magnitudes
of extinction in the observed frame $i$ band:
\begin{equation}
(\Delta z)_{j,dust}  = \sum_{k, m_k+A_j<19.1} \Delta X_k \times P(z_{abs,j}, 
W_j, m_k+A_j)
\end{equation}
where the sum is over all quasars brighter than our survey magnitude
limit, $k$ denotes a quantity associated with the background quasar
and $j$ a quantity associated with the absorber to be found. $\Delta
X$ is the redshift path defined by the limits of the survey given in
Section \ref{sample_detail}:
\begin{equation} 
\Delta X = min[1.3,z_{qso}-0.1]- max[0.4,\frac{1250(1+z_{qso})}{2796} -1]
\end{equation}

The probability, $P$, of detection of a line of given equivalent width and
redshift, in a spectrum of a quasar of given magnitude, is calculated from Monte
Carlo simulations:  Gaussian line features are placed in the quasar spectra, our
initial detection algorithm is run and the rate of recovery of the lines is
measured.  Including the probability of detection of the
\mgii\,$\lambda\lambda$2796,2803 doublets has a negligible effect on the final
results, due to their being considerably stronger than the \caii\ lines (mean
equivalent width of $W_{\lambda2796}=2.16\,$\AA).

The extinction in individual absorbers is estimated in the same way as
for the subsamples, except that the absorber spectra are not combined
into composites before the extinction curve is fitted. While the error
on each individual reddening measure is large due to the unknown
colour of the background quasar, the observed trends with equivalent
width of \caii\ give us confidence that on average the correct value
is measured.

A correction factor, $\Delta z_{no-dust}/\Delta z_{dust}$, is calculated for
each observed absorption line system, which tells us the number of absorbers
missed for each one seen.  An estimate is thus obtained of the total number of
absorbers missing from our sample.  Of course, this method only allows us to
estimate correction factors for regions of absorber parameter space which are
well sampled by our observations.  At high reddening values the number of
absorbers drops dramatically and we are unable to correct reliably for absorbers
with an \EBV$>0.2$ and $>0.25$ for the low- and high-redshift samples
respectively.  Our results are therefore only applicable to absorbers with
reddening values below these limits.  To place these \EBV\ limits into context,
the reddening present in galaxies at $z\sim2$, selected via ultraviolet
continuum flux, has a median of 0.15\,mag and ranges between 0.0 and 0.4\,mag
\citep{2005ApJ...626..698S}.

\end{appendix}

\end{document}